\documentclass{appolb}
\usepackage{graphicx}
\usepackage{amsmath}

\begin{document}
\title{A Study  of Charge Radii and Neutron Skin Thickness  near Nuclear Drip Lines%
}
\author{Virender Thakur\thanks{virenthakur123@gmail.com}, Shashi K Dhiman\thanks{shashi.dhiman@gmail.com}
\address{Department of Physics, Himachal Pradesh University, Summer-Hill, Shimla-171005, INDIA}
}
\maketitle
\begin{abstract}
 We studied the charge radius, rms radius and neutron skin thickness $\Delta r_{np}$ in even-even isotopes of Si, S, Ar and Ca and isotones of N =20, 28, 50 and 82. The $\Delta r_{np}$ in doubly-magic $^{48}$Ca, $^{68}$Ni, $^{120,132}$Sn 
and $^{208}$Pb nuclei has also been calculated. Theoretical calculations are done with the Hartree-Fock-Bogoliubov theory with the effective Skyrme interactions. Calculated theoretical estimates are in good agreement with the recently available experimental data.
The charge radii for Si, S, Ar and Ca isotopes is observed to be minimum at neutron number N =14.
The theoretically computed results with UNEDF0 model parameterization of functional are reasonably reproducing the experimental data for $\Delta r_{np}$ in $^{48}$Ca, $^{68}$Ni and $^{120,132}$Sn. The  energy density functional of UNEDF1 model provides much improved result of $\Delta r_{np}$ for $^{208}$Pb.
\end{abstract}
  
\section{Introduction}
\label{intro}
Appreciable experimental progress in producing and analyzing exotic nuclei \cite{Lunderberg2016,Gade2009} has brought renaissance of nuclear structure models. In recent years \cite{Coraggio2009,Papakonstantinou2018,Niksic2008,Bhuyan2018}, The production of more and more new isotopes has revived the interest in nuclear structure models. In nuclear physics, understanding the structure of the atomic nucleus is one of the key challenges.  The study of those nuclei which are lying far from the line of $\beta$-stability play a crucial role in the understanding of nuclear physics. The limit of nuclear existence is reached while going far away from stability line. An unbound nuclear system disintegrates quasi-
instantaneously (in a time interval of the order of $10^{-21}$s. Points on
the nuclear chart which corresponds to nuclei that are unbound to proton or neutron emissions in the
ground state is called the proton or neutron drip line. These drip lines form the edges
of the nuclear chart. About 280 stable  nuclear species are found on
 Earth but, according to latest estimates, from 5000 to 7000 bound nuclei should
exist in the Universe. Till today, about 2000 in numbers have been synthesized and observed. Very
little information is available about these unstable nuclei.\\
Even though the number of undiscovered bound nuclei is very large but we are able to
make single big steps by studying a few of specific nuclei. These nuclei act as milestones
in setting  new effects that arise in extreme conditions of isospin asymmetry. We have chosen Si, S, Ca and Ar nuclei for our interest as lot of research has been done on these nuclei recently \cite{Gade2006,Utsuno2012,Tostevin2013,Bhattacharyya2008,Utsuno2015,Stroberg2014}. The microscopic structure of these nuclei is
of particular interest for the field of  astrophysics: the neutron-rich N$\simeq$28 nuclei play an important
role in the nucleosynthesis of the heavy Ca-Ti-Cr isotopes \cite{Sorlin1993}. As these nuclides also become
experimentally accessible \cite{Sorlin1993,Schneider1994}, they can provide a testing ground for studying exotic
nuclei.\\ Exotic nuclei, particularly those near the drip lines, are at the core of one of the fundamental
questions driving nuclear structure and astrophysics today: What are the limits of nuclear binding? These  Exotic nuclei
play a critical role in both informing theoretical models as well as in our understanding of the origin of the heavy
elements. Of the nuclei considered here, those with N = Z ($^{28}$Si, $^{32}$S, $^{36}$Ar and $^{40}$Ca) have
been studied extensively both theoretically and experimentally. Among different theoretical approaches used are the shell-correction approach \cite{Leander1975}, Hartree-Fock \cite{Jaqaman1984,Flocard1984} 
and Hartree-Fock-Bogolyubov methods \cite{Girod1983}, relativistic mean field \cite{Furnstahl1987,Fink1989,Patra1993,Zhang1994}, shell
model \cite{Carchidi1986,Zhang1994(1)}, $\alpha$-cluster model \cite{Zhang1994(1)} and others.\\ To throw some light on the physics of exotic neutron-rich and magic-nuclei near drip lines, we
performed calculations based on the self-consistent mean-field theories: Hartree-Fock-Bogoliubov
(HFB) with the Skyrme effective interactions approach. The physical observables of interest are nuclear root mean square radius (R$_{rms}$), nuclear charge radius ($R_{c}$) and neutron skin thickness
($\Delta r_{np}$). The nuclear charge radius is one of the most fundamental nuclear properties that describes the effect of effective interactions on nuclear structure \cite{Angeli2013,Yang2018}.  It can be measured experimentally by methods based on the electromagnetic interaction between the nucleus and electrons or muons. One of the parameter that can be expressed through the nucleon density distribution is neutron skin thickness which depends on the properties of the nuclear surface. For large neutron excess, the bulk of neutron density is believed to extend beyond the proton density creating a sort of 
 "neutron skin". To determine the nature of the neutron distribution accurately in a nuclei has received considerable attention 
 in recent years \cite{Centelles2009,Brown2017,Brown2009}. As neutron number increases, the radius of the neutron density distribution becomes larger than that of the protons, reflecting the pressure of the symmetry energy.\\
\section{Theoretical Framework}
The present study has been done by using the model based on  Hartree-Fock-Bogoliubov Theory \cite{Ring1980,Bender2003}. A brief discussion of the theory is given below.\\
In this theory, the pairing field and the mean field are given equal footings. Zero range Skyrme interactions are used in the self consistent mean field part \cite{Skyrme1959}. For the Skyrme forces, The HFB Energy has the form of local energy density functional \cite{Stoitsov2005,Stoitsov2013}:
\begin{equation}
\label{eq:eqA}
   E[\rho,\tilde{\rho}] = \int d^{3}r {\cal H}(r)
  \end{equation}
 where
 \begin{equation}
 \label{eq:eqB}
  {\cal H}(r) = H(r) + \tilde{H}(r)
 \end{equation}
 is the sum of the mean-field and paring energy densities.
 The wave function $\varPhi$ depends upon the density matrix $\rho$ and the pairing tensor $\kappa$ and in terms of these two parameters, the HFB energy takes the form \cite{Stoitsov2005}:
 \begin{equation}
 \label{eq:eqC}
 E[\rho,\kappa] = \frac{\langle\phi|H|\phi\rangle}{\langle\phi|\phi\rangle} = Tr\left[\left(e + \frac{1}{2}\Gamma\right)\rho\right] - \frac{1}{2}Tr[\Delta{\kappa}^{*}]
 \end{equation}
 where $\Gamma$ is Hartree-Fock potential and $\Delta$ stands for the pairing potential. They are defined in their explicit form as
 \begin{equation}
 \label{eq:eqD}
 \Gamma_{n_{1}n_{3}} = \sum_{n_{2}n_{4}}\overline{v}_{n_{1}n_{2}n_{3}n_{4}\rho_{n_{4}n_{2}}},\quad\Delta_{n_{1}n_{2}} = \frac{1}{2}\sum_{n_{3}n_{4}}\overline{v}_{n_{1}n_{2}n_{3}n_{4}}\kappa_{n_{3}n_{4}}
 \end{equation}
 
This leads to the HFB Hamiltonian matrix \cite{Stoitsov2005}:
 \begin{eqnarray}
 \label{eq:eqE}
 \left(\begin{array}{cc}h^{(q_{k})} - \lambda^{(q_{k})}& \tilde{h}^{(q_{k})}\\
        \tilde{h}^{(q_{k})}& -h^{(q_{k})} + \lambda^{(q_{k})}
       \end{array}\right)\left(\begin{array}{c}U_{k}\\V_{k}\end{array}\right)\nonumber\\ = E_{k}\left(\begin{array}{c}U_{k}\\V_{k}\end{array}\right)
\end{eqnarray}
where $E_{k}$ denotes the quasiparticle energies, and $\lambda^{q_{k}}$ represent chemical potential. The matrices
\begin{equation}
\label{eq:eqE}
 h_{\alpha\beta}^{(q)} = \langle\varPhi_{\alpha}|h_{q}|\varPhi_{\beta}\rangle, \qquad \tilde{h}_{\alpha\beta}^{(q)} = \langle\varPhi_{\alpha}|\tilde{h}_{q}|\varPhi_{\beta}\rangle
\end{equation}
are defined for a given proton $(q_{k} = +1/2)$ or neutron = $(q_{k} = -1/2)$ block \cite{Stoitsov2005}.
\section{Results And Discussions}
We present our results for root mean square radii and charge radii reflecting size of the nucleus. Neutron skin thickness for  isotopes of Si, S, Ar and Ca with 
particular reference to the disappearance of magic number N = 28 with Z = 14 in $^{42}$Si \cite{Bastin2007} nuclei has also been analysed. Beside this,  
the shell structure evolution, nuclear radii and neutron skin thickness in the 
isotones at N = 20, 28, 50 and 82 and doubly magic nuclei in $^{48}$Ca, $^{68,78}$Ni, $^{120,132}$Sn and $^{208}$Pb have also been investigated.
The theoretical results are computed from Axially deformed solution of the
Skyrme-Hartree-Fock-Bogoliubov equations using
the harmonic oscillator basis \cite{Stoitsov2005,Stoitsov2013}, where the mean field potentials are constructed with UNEDF0  and UNEDF1 parameterizations \cite{Kortelainen2010,Korte2012}. 
\subsection{Size of the Nucleus}
Nuclear radii of proton and neutron distributions are the key observables to determine the size of the nuclear system. Even, these observables
can be directly related to fundamental bulk properties of nuclear matter and also to the nature of nuclear 
interactions \cite{Reinhard2016,Hagen2015}. In exotic neutron-rich and heavy nuclei the excess of 
neutrons give rise to a neutron skin t the size of a nuclear system such as; nuclear charge radii, neutron distributions and
neutron skin thickness have been found to correlate empirically with a number of observables in finite nuclei \cite{Angeli2013}
to neutron matter \cite{Reinhard2016,Hagen2015,Agrawal2006}. Hence, it further beautifully links with a plausible constraints found 
to impose on the equations of state high density matter in neutron stars \cite{Lattimer2013}.

The goal of this section is to understand the relations between nuclear charge and neutron radii, and compare our theoretical
results with available experimental data.\\

\begin{figure}
\centering
\includegraphics[trim=0 0 0 0,clip,scale=0.5]{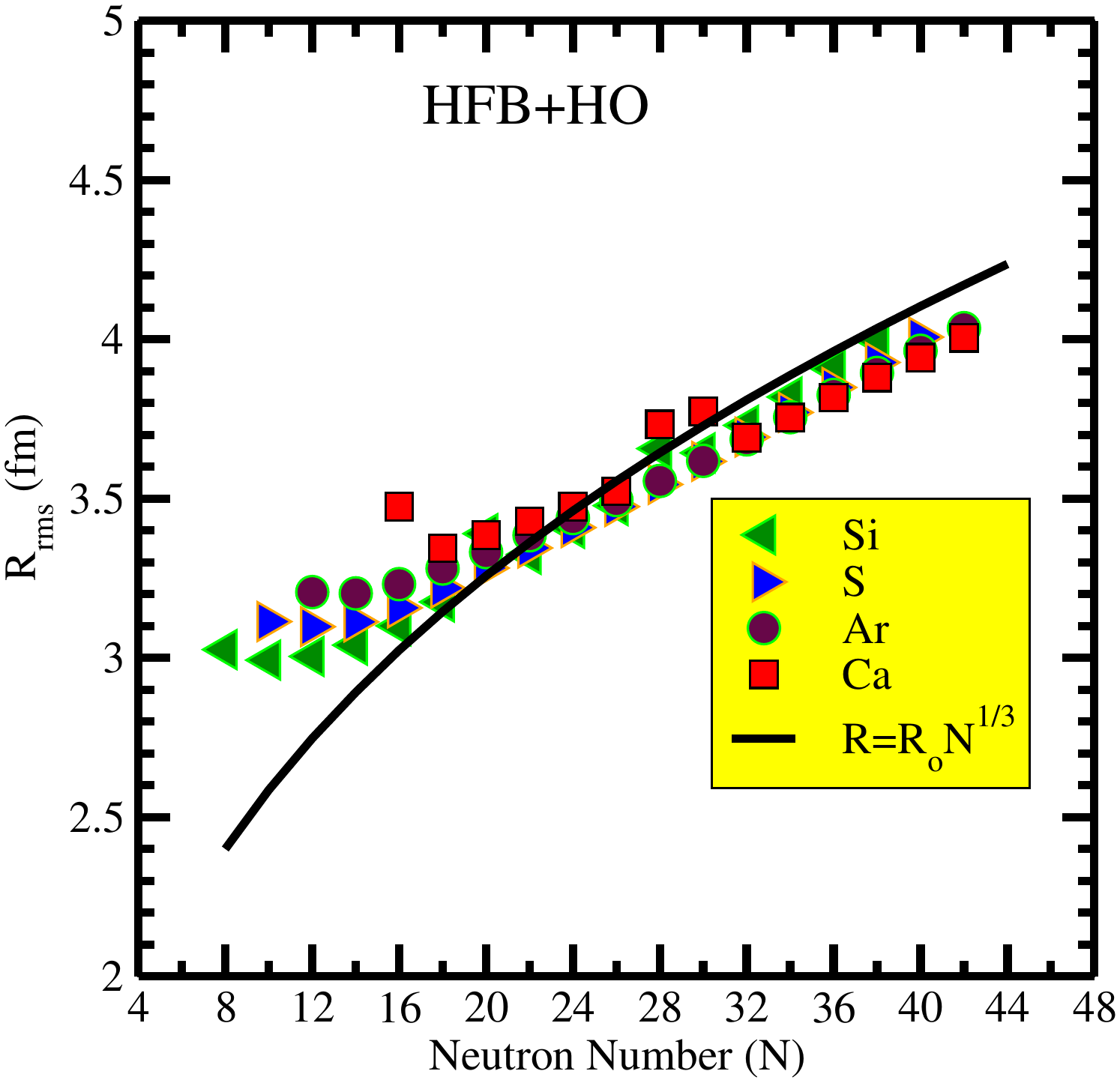} 
\caption{\label{fig1}(color online)  The root mean square radius in fm plotted as a function of Neutron
Number (N), for the exotic nuclei of Silicon, Sulphur, Argon and Calcium. The theoretical estimates are computed from Axially deformed solution of the
Skyrme-Hartree-Fock-Bogoliubov equations using
the harmonic oscillator basis (HFB+HO), based on Energy Density Functional (EDF) parameterization
UNEDF0.}
\end{figure}

\begin{figure}
\centering
\includegraphics[trim=0 0 0 0,clip,scale=0.35]{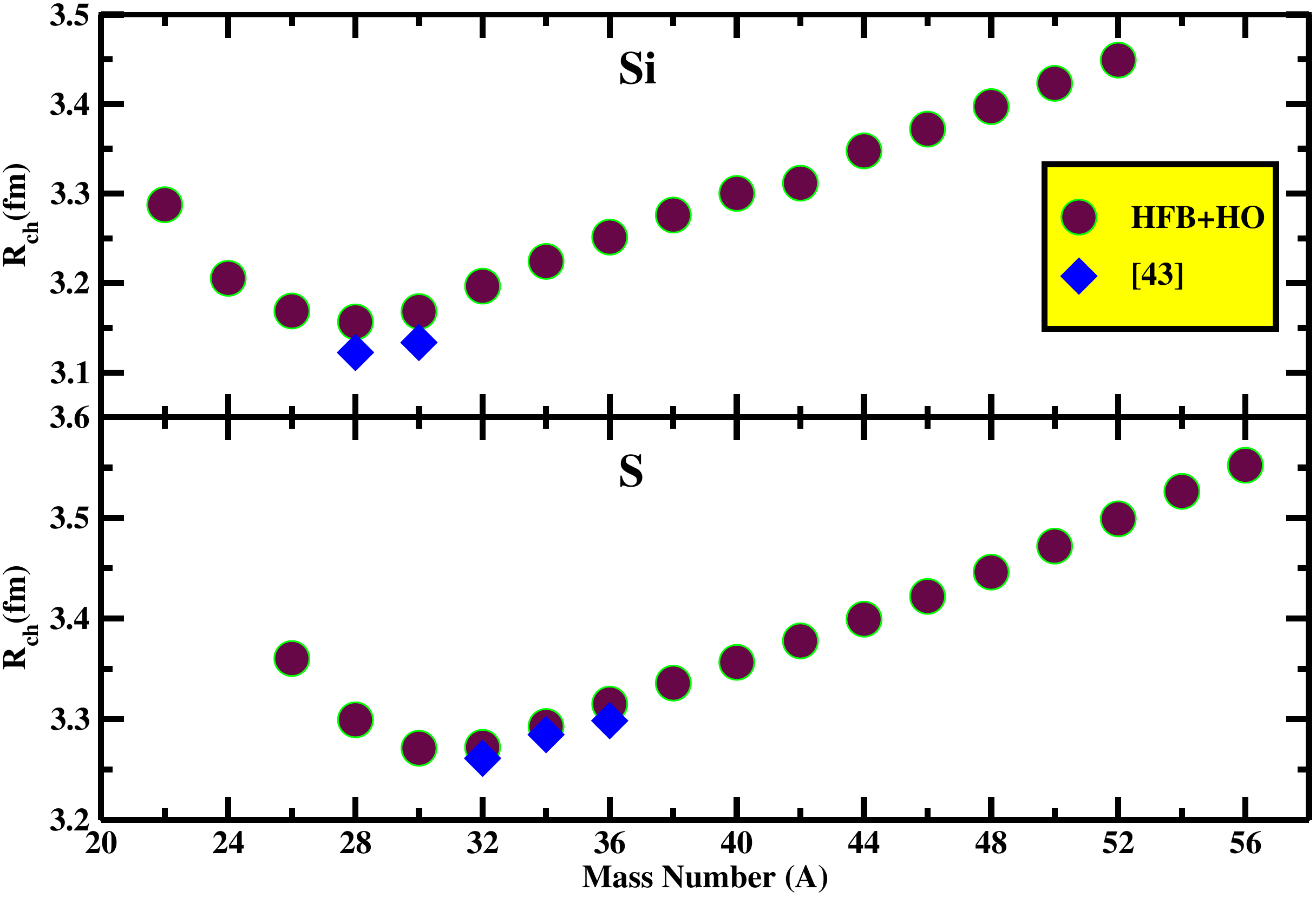} 
\caption{\label{fig2}(color online)  The charge radius in fm plotted as a function of Mass Number (A), 
for the exotic nuclei of Silicon (upper panel) and Sulphur (lower panel). The theoretical estimates are computed from Axially deformed solution of the
Skyrme-Hartree-Fock-Bogoliubov equations using
the harmonic oscillator basis (HFB+HO), based on Energy Density Functional (EDF) parameterization
UNEDF0. The experimental
data is taken from \cite{Angeli2013}.}
\end{figure}

\begin{figure}
\centering
\includegraphics[trim=0 0 0 0,clip,scale=0.35]{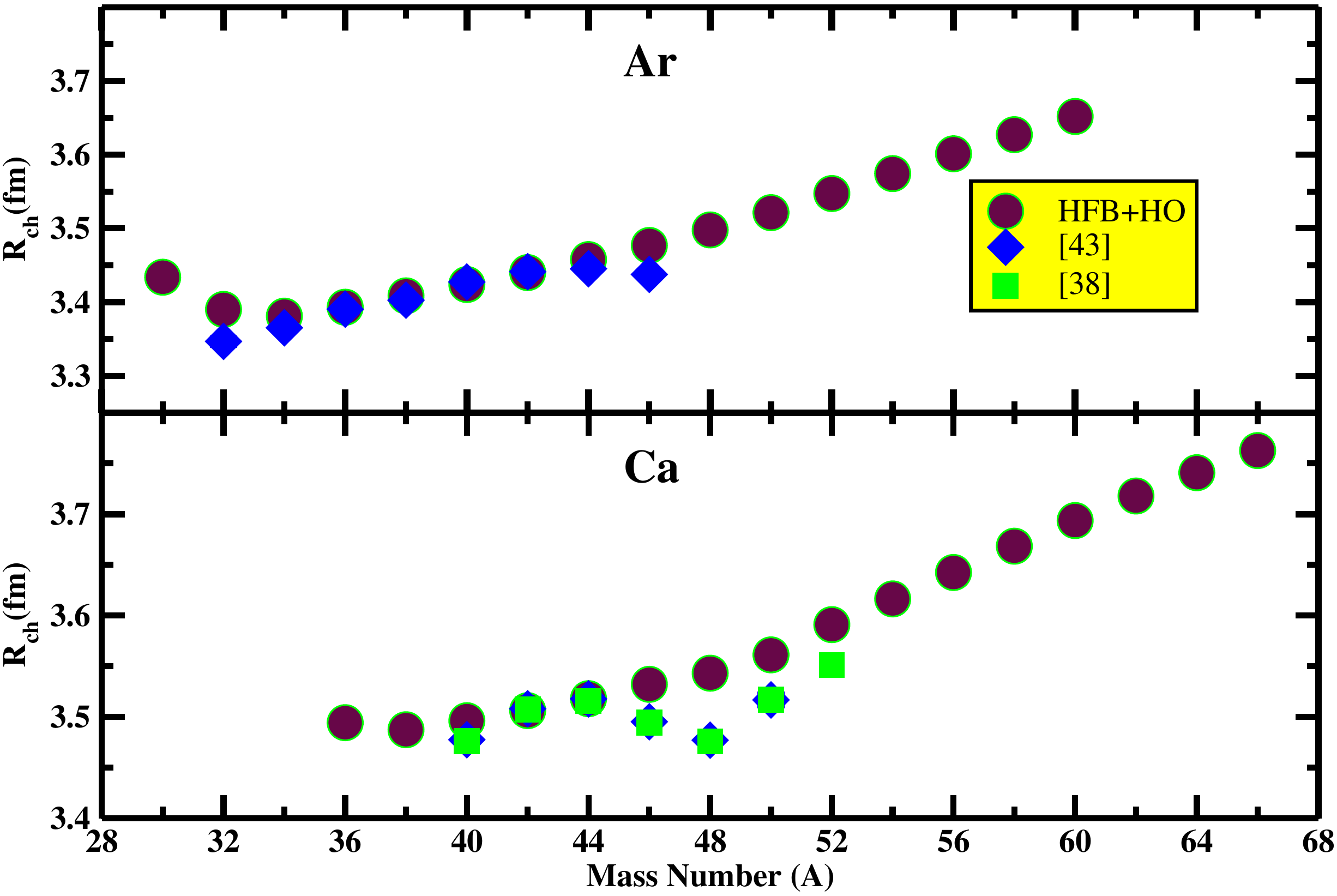} 
\caption{\label{fig3}(color online)  The charge radius in fm plotted as a function of Mass Number (A), 
for the exotic nuclei of Argon (upper panel) and Calcium (lower panel). The theoretical estimates are computed from Axially deformed solution of the
Skyrme-Hartree-Fock-Bogoliubov equations using
the harmonic oscillator basis (HFB+HO), based on Energy Density Functional (EDF) parameterization
UNEDF0. The experimental
data is taken from \cite{Angeli2013,Garcia2016}.}
\end{figure}

Here in Fig. (\ref{fig1}), results for the root mean square radius $R_{rms}$ in fm plotted as a function of nuclear neutron number N, for the even-even  
exotic to stable nuclides of  Si (green4 left triangles), 
Sulphur (blue right triangles) Ar (maroon circles) and Ca (red squares). 
It can be seen from Fig. (\ref{fig1}) that the RMS radii of the proton-rich or neutron-rich lie away from the solid line of radii 
 of stable nuclear systems. In case of proton rich nuclei, the values of RMS radii are smaller due to the smaller neutron
 distributions as compared to neutron rich nuclei. That is  why, the theoretically calculated points showing RMS radius for exotic even-even nuclei of Si, S, Ar and 
 Ca for proton rich nuclei as shown in Fig. (\ref{fig1})  lies little above
 from the stability curve. But, for the neutron rich nuclei the theoretically calculated points for the RMS radius of all the four nuclei (Si, S, Ar and Ca) lies little lower than the stability curve. 
 The solid line here represents the stability curve for the given mass 
 region of nuclides which is computed by using formula $R =R_{0} N^{1/3}$.\\
 The results for the charge radii $R_{ch}$ in fermi meter as a function of nuclear mass number A for the nuclides of Argon, 
Calcium, Silicon and Sulphur are shown in Figs. (\ref{fig2} and \ref{fig3}). The theoretical charge radius is calculated using the formulae \cite{Paar} :
\begin{equation}
\label{eq:Rch}
 R_{ch} = \sqrt{r_{p}^{2} + 0.64} \hspace{0.2cm} fm
\end{equation}
where, $r_{p}$ denotes the rms radius of the proton density distribution and term $0.64 \hspace{0.2cm} {fm}^2$ accounts for the finite size of proton. The experimental measurements of $R_{ch}$ are also shown for comparison and represented by symbol diamond with error bars \cite{Angeli2013}.

\begin{figure}
\centering
\includegraphics[trim=0 0 0 0,clip,scale=0.35]{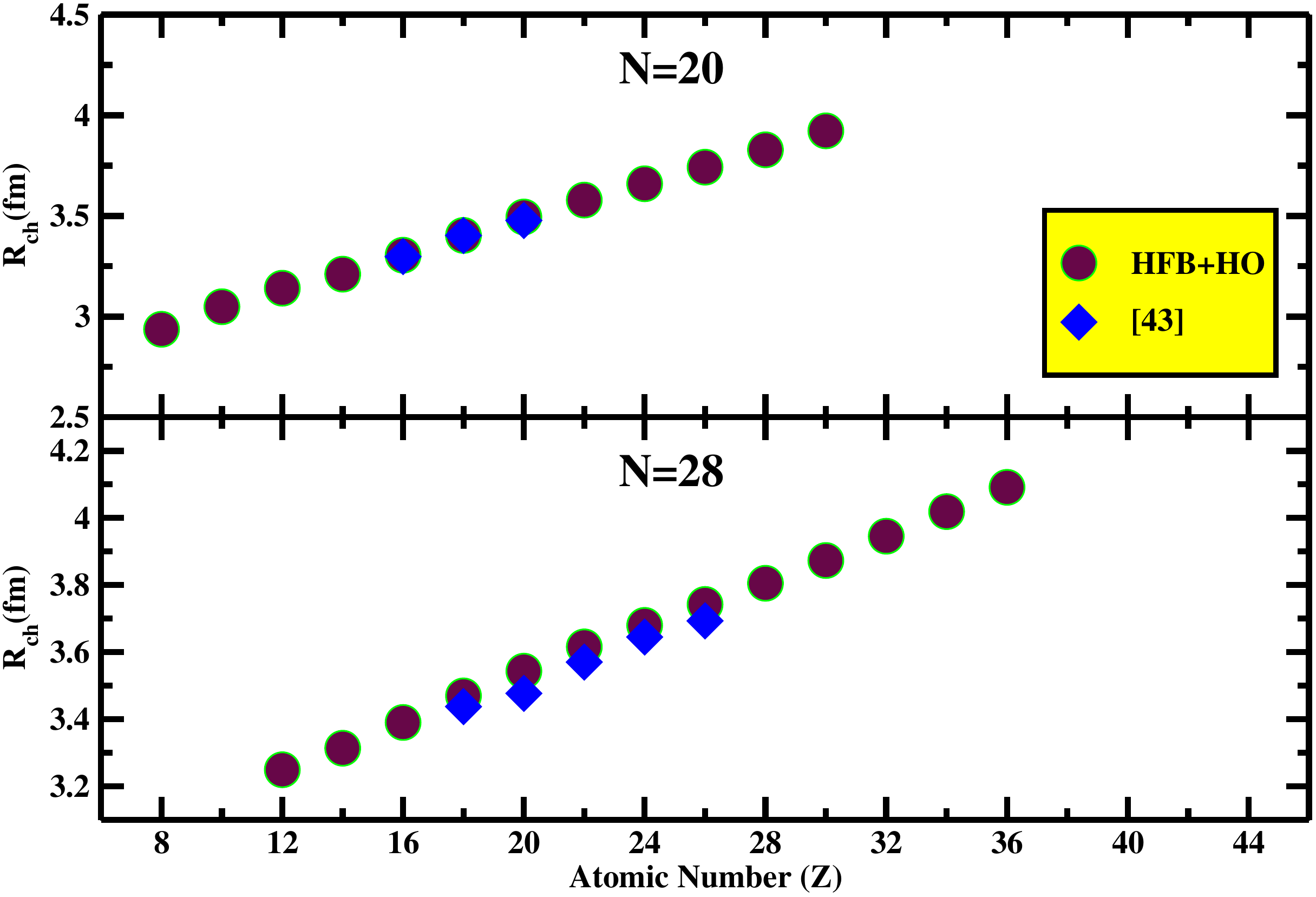} 
\caption{\label{fig4}(color online)  The charge radius in fm plotted as a function of  Atomic Number (Z), 
for the exotic nuclei of $N = 20$ (upper panel) and $N = 28$ (lower panel) isotones. The theoretical estimates are computed from Axially deformed solution of the
Skyrme-Hartree-Fock-Bogoliubov equations using
the harmonic oscillator basis (HFB+HO), based on Energy Density Functional (EDF) parameterization
UNEDF0. The experimental
data is taken from \cite{Angeli2013}.}
\end{figure}

\begin{figure}
\centering
\includegraphics[trim=0 0 0 0,clip,scale=0.35]{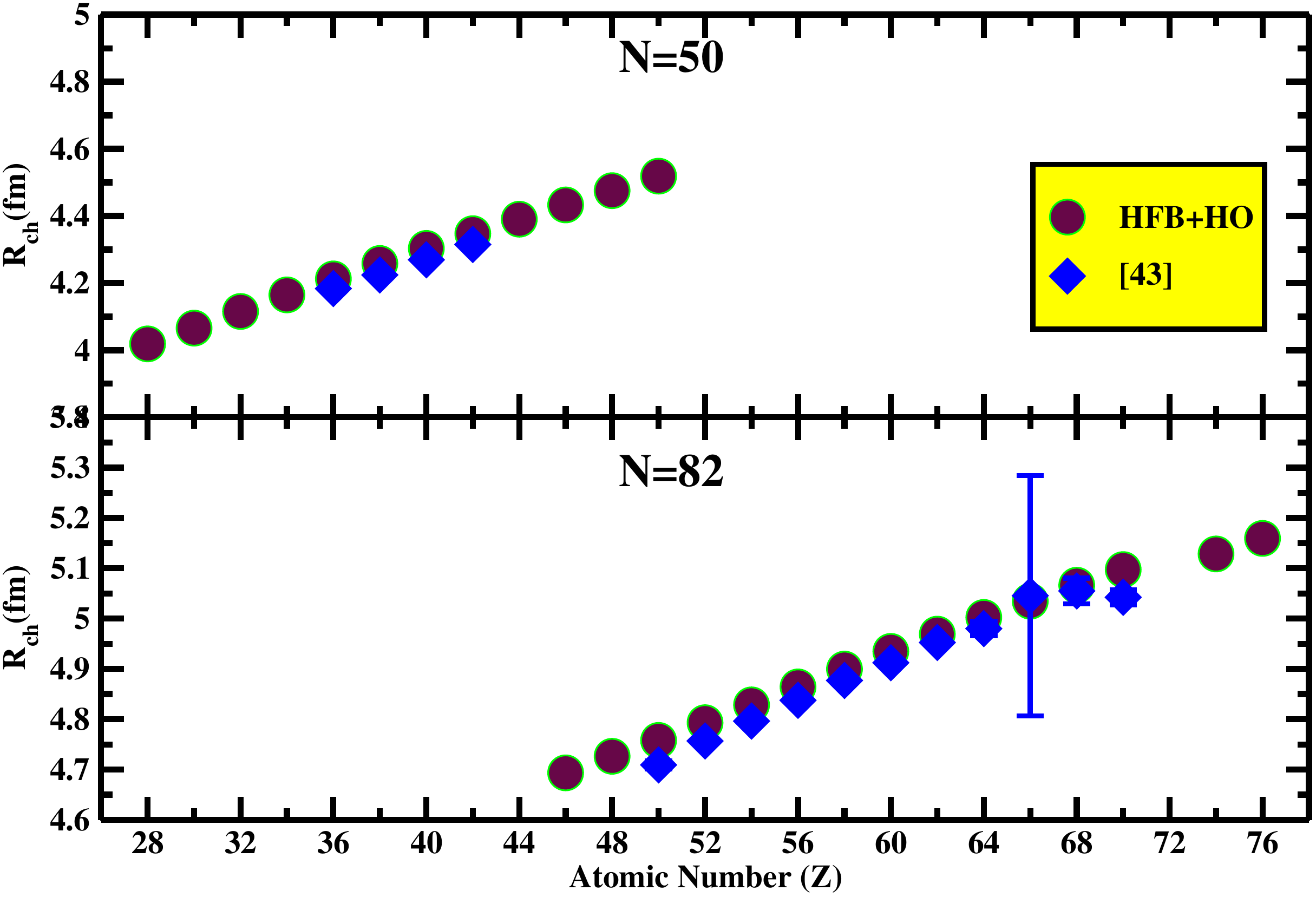} 
\caption{\label{fig5}(color online)  The charge radius in fm plotted as a function of Atomic Number (Z), 
for the exotic nuclei of $N = 50$ (upper panel) and $N = 82$ (lower panel) isotones. The theoretical estimates are computed from Axially deformed solution of the
Skyrme-Hartree-Fock-Bogoliubov equations using
the harmonic oscillator basis  (HFB+HO), based on Energy Density Functional (EDF) parameterization
UNEDF0. The experimental
data is taken from \cite{Angeli2013}.}
\end{figure}

It is clear from Figures that, those  nuclei which lie in the proton drip line and 
 the neutron drip line have larger charge radius in comparison to the other stable nuclei. The value of the charge radius is found to be  minimum for 
 $^{28}$Si nucleus with N = Z = 14 as shown in upper panel of Fig. (\ref{fig2}), $^{30}$S nucleus with N = 14 and Z = 16 in lower panel of Fig. (\ref{fig2}), $^{32}$Ar with N = 14 
and Z = 18 in upper panel of Fig., indicating sub-shell closure with $d_{5/2}$ occupying neutron N or proton Z = 14, as recently investigated Z = 14
sub-shell closure in 
$^{42}$Si nucleus \cite{Fridmann2005}. 
The theoretical results are producing the experimental data of  charge radii $R_{ch}$ in isotopes in Si, S, Ar and Ca with
the UNEDF0 parameterizations \cite{Kortelainen2010}.
Similarly, the variation of charge radii $R_{ch}$ in fm as a function of nuclear mass number for various isotones of $N = 20,28,50$ and $82$ against Atomic Number (Z) 
are displayed in Figs. (\ref{fig4} and \ref{fig5}).
The experimental results are also shown in Figures for comparison with theoretical values. It can be observed from the 
Figures that  EDF parameterizations (UNEDF0) are reproducing experimental data of charge radii.
In case of N = 20 isotones, the experimental data is available for nuclides with A = 36 to A = 40 ($^{36}_{16}S$, $^{38}_{18}{Ar}$ and $^{40}_{20}{Ca}$) as shown
in upper panel of Fig. For N = 28,the experimental data is available for nuclides  with A = 46 to A = 54 ($^{46}_{18}{Ar}$,
$^{48}_{20}{Ca}$, $^{50}_{22}{Ti}$ ,$^{52}_{24}{Cr}$
and $^{54}_{26}{Fe}$) as shown in lower panel of Fig. (\ref{fig5})
In case of  N = 50 isotones, the experimental data taken from ref    \cite{Angeli2013}  is available for for the nuclides with mass number A = 86 to A = 92 ($^{86}_{36}{Kr}$, $^{88}_{38}{Sr}$, $^{90}_{40}{Zr}$ and $^{92}_{42}{Mo}$) as
shown in upper panel of Fig. (\ref{fig5}) However for N = 82, we have experimental results for A = 132 to A = 152 
($^{132}_{50}{Sn}$, $^{134}_{52}{Te}$, $^{136}_{54}{Xe}$, $^{138}_{56}{Ba}$, $^{140}_{58}{Ce}$, $^{142}_{60}{Nd}$, $^{144}_{62}{Sm}$,
$^{146}_{64}{Gd}$, $^{148}_{66}{Dy}$, $^{150}_{68}{Er}$ and $^{152}_{70}{Yb}$) as shown in lower panel of Fig. (\ref{fig5}),
and our theoretical estimates are in good agreement with these experimental results.

\subsection{Neutron Skin Thickness} 
\begin{figure}
\centering
\includegraphics[trim=0 0 0 0,clip,scale=0.5]{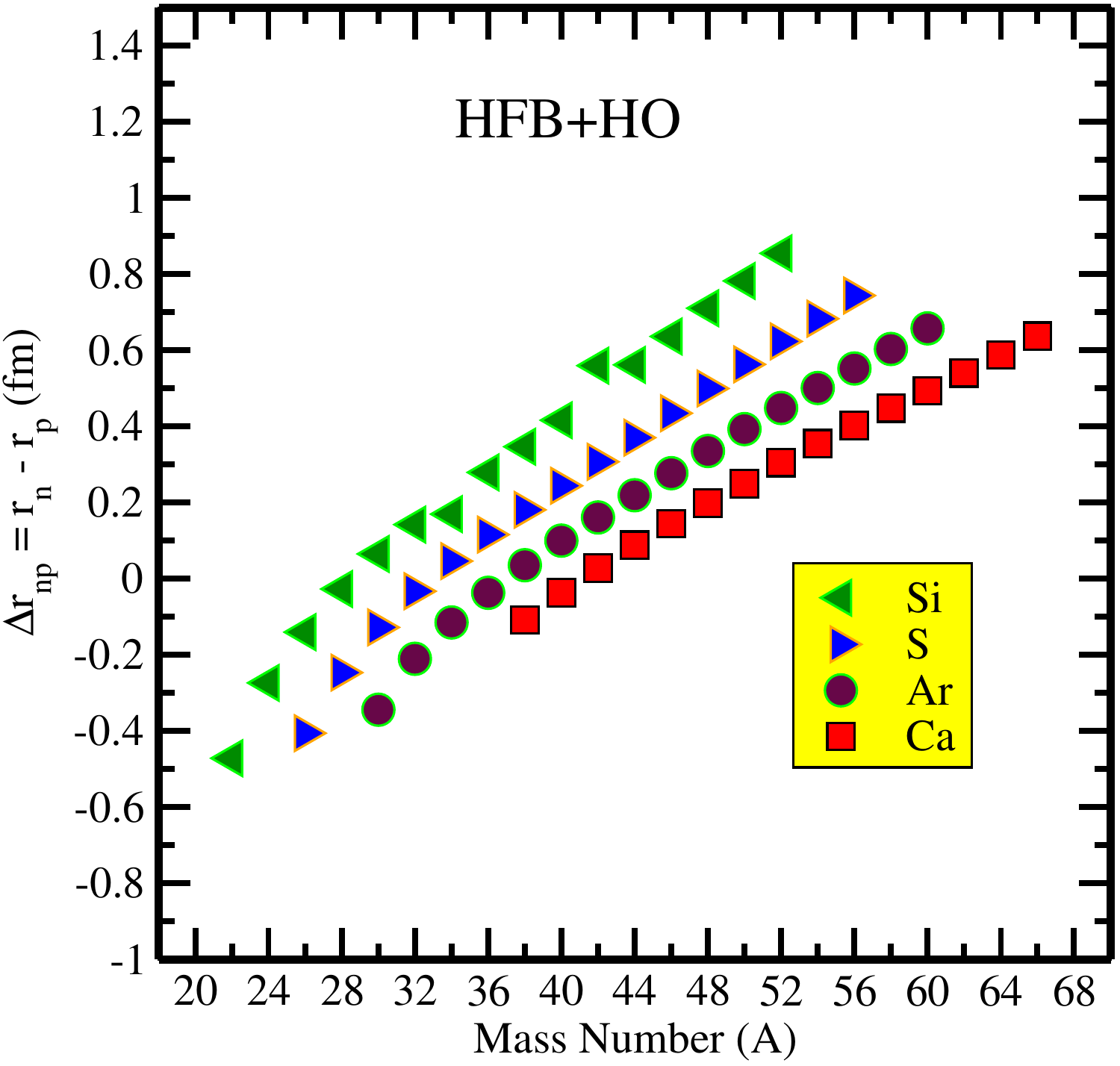}
\caption{\label{fig6}(color online)  The neutron skin thickness ${\Delta r_{np} = r_n - r_p}$ in fm, plotted as a function of Mass Number (A),
for the chain of isotopes of Silicon, Sulphur, Argon and Calcium. The theoretical estimates are computed from Axially deformed solution of the
Skyrme-Hartree-Fock-Bogoliubov equations using
the  harmonic oscillator basis  (HFB+HO), based on Energy Density Functional (EDF) parameterization
UNEDF0.}
\end{figure}

\begin{figure}
\centering
\includegraphics[trim=0 0 0 0,clip,scale=0.35]{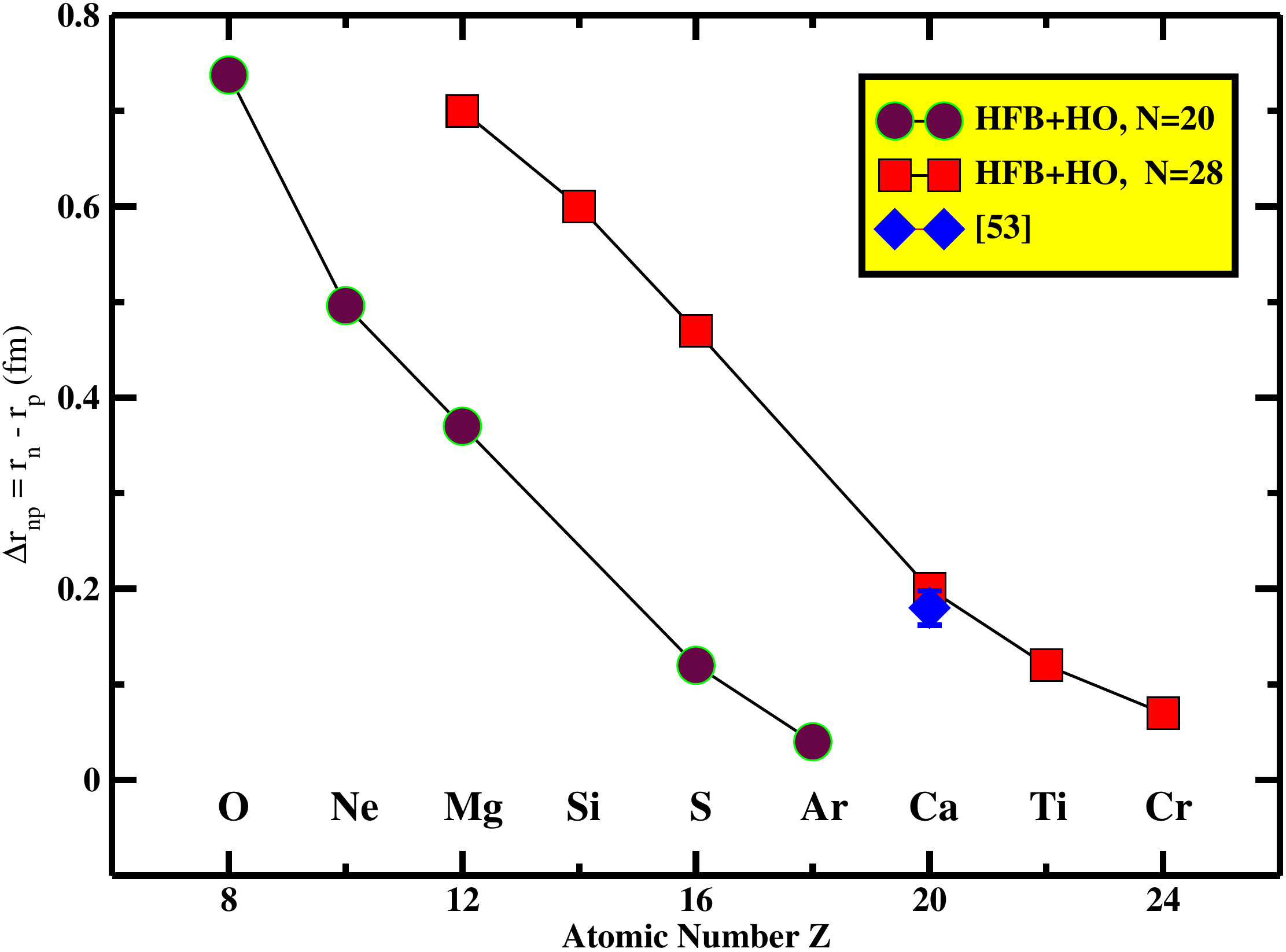}
\caption{\label{fig7}(color online)  The neutron skin thickness ${\Delta r_{np} = r_n - r_p}$ in fm, plotted as a function of
Atomic Number (Z),
for the isotonic chain of of $N = 20$ and $N = 28$. The theoretical estimates are computed from Axially deformed solution of the
Skyrme-Hartree-Fock-Bogoliubov equations using
the harmonic oscillator basis (HFB+HO), based on Energy Density Functional (EDF) parameterization
UNEDF0.}
\end{figure}

\begin{figure}
\centering
\includegraphics[trim=0 0 0 0,clip,scale=0.35]{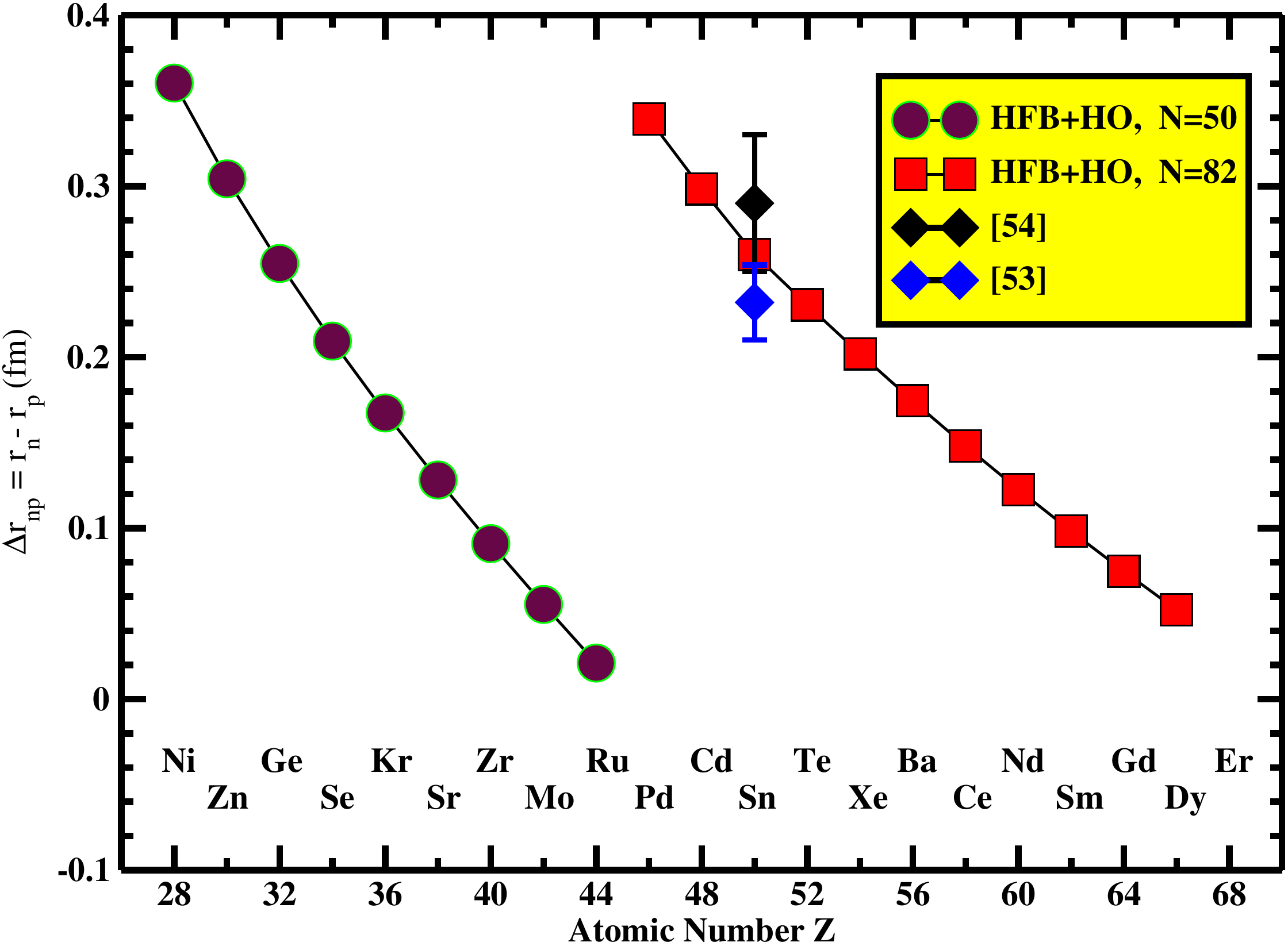}
\caption{\label{fig8}(color online) The neutron skin thickness ${\Delta r_{np} = r_n - r_p}$ in fm, plotted as a function
of Atomic Number (Z),
for the isotonic chain of of $N = 50$ and $N = 82$. The theoretical estimates are computed from Axially deformed solution of the
Skyrme-Hartree-Fock-Bogoliubov equations using
the  harmonic oscillator basis (HFB+HO), based on Energy Density Functional (EDF) parameterization
UNEDF0.}
\end{figure}

The neutron skin thickness $\Delta r_{np}$ is defined as the difference between the nuclear rms radii obtained using the density 
distributions for point neutrons and point protons, i. e., 
\begin{equation}
\label{eq:rnp}
\Delta r_{np} = \sqrt{r^2_n} - \sqrt{r^2_p},
\end{equation}
 where $r_n$ refer to  the neutron rms radius and $r_p$ destine the proton rms radius. 
 It is well understood from the ref. \cite{Agrawal2006} that the accurate measurement of the neutron skin thickness would place rigid  constraints on the density
 dependence of the nuclear symmetry energy denoted as $S(\rho)$.  The density dependence of the nuclear symmetry energy have direct consequences 
 in finite nuclear matter  and nuclear dense matter of astrophysical interest. This is why, it is investigated extensively from both 
 theoretical and experimental perspectives. Within available experimental technology in nuclear physics, the nuclear symmetry energy 
 can not be measured directly 
 but the information of this fundamental quantity can be extracted from the neutron skin thickness and electric dipole 
 polarizability \cite{Reinhard0000}. 
 Whereas, the direct measurement of neutron distribution is also  extremely difficult, therefore, the recent experimental probes have focused 
 on the accurate measurements of electric dipole polarizability $\alpha_D$ in $^{208}Pb$ \cite{Tamii2011}, $^{120}Sn$ \cite{Hashimoto2015}
 and $^{68}Ni$ \cite{Rossi2013},
 while for the extraction of $\alpha_{D}$ in $^{48}Ca$  by the Darmstadt-Osaka collaboration is working on at 
  Jefferson Laboratory \cite{Nazarewicz2014}. These nuclei are very interesting from point of view of nuclear structure physics, because they are 
  neutron-rich and doubly magic nuclear systems, can be understood with recently developed {\it ab-initio} nuclear  density functionals theory. 
  The studies of these nuclear structure observable can provide a constraint for the size of neutrons star \cite{Hagen2015}.\\
  In the Fig. (\ref{fig6}), we present the neutron skin thickness ${\Delta r_{np} = r_n - r_p}$ in fm, plotted as a function of nuclear mass
number A, for the chain of even-even exotic isotopes of Silicon (green4 left triangles), Sulphur (blue right triangles), Argon (maroon circles) and Calcium (red squares).  The theoretical estimates are computed from Axially deformed solution of the
Skyrme-Hartree-Fock-Bogoliubov equations using
the  harmonic oscillator basis model (HFB+HO), based on EDF parameterization UNEDF0. 
The magnitudes of skin thickness is increasing systematically with increase in the neutron number in the isotopes in all the elements.\\
In case of isotones, whose results are shown in Figs. (\ref{fig7} and
\ref{fig8}), we find that there is decrease in the neutron skin depth with increase in atomic number (Z) which
signifies that with increase in proton number, there is decline in neutron density. This is because of the fact that as the proton number raises, 
the $r_p$ value also increases while the neutron density is unaffected. We have taken in account the isotonic chains of $N =20$, $N = 28$.

Fig. (\ref{fig7}) presents the neutron skin thickness in isotones corresponding to neutron shell closures 
at magic N = 20 (upper panel) and N = 28 (lower panel). In upper panel of Fig. (\ref{fig7}), we present the neutron skin thickness ${\Delta r_{np} = r_n - r_p}$ in fm,
plotted as a function of atomic number Z, for the even-even isotonic chain of N = 20, for $^{28}$O, $^{30}$Ne, $^{32}$Mg, $^{34}$Si, 
$^{36}$S, and $^{38}$Ar nuclei.
It can be seen that the magnitude of the skin thickness is decreasing as on  increasesing the proton numbers except for 
$^{34}$Si with Z = 14, where it rise a little bit from the previous nuclei ($^{32}$Mg) as calculated from HFB+HO model with 
Parameterization UNEDF0. Neutron skin thickness 
varies from $0.10 - 0.7$ fm in magnitude. The $^{28}$O nucleus has large  ${\Delta r_{np}}$ magnitudes of about $0.7$ fm  computed with HFB+HO model 
due to the larger expected neutron distributions than proton numbers. Whereas, in case of $^{42}$Ti nucleus where Z = 22 and N = 20,
the magnitude of the ${\Delta r_{np}} = 0.10$ fm, is very small.

\begin{figure}
\centering
\includegraphics[trim=0 0 0 0,clip,scale=0.4]{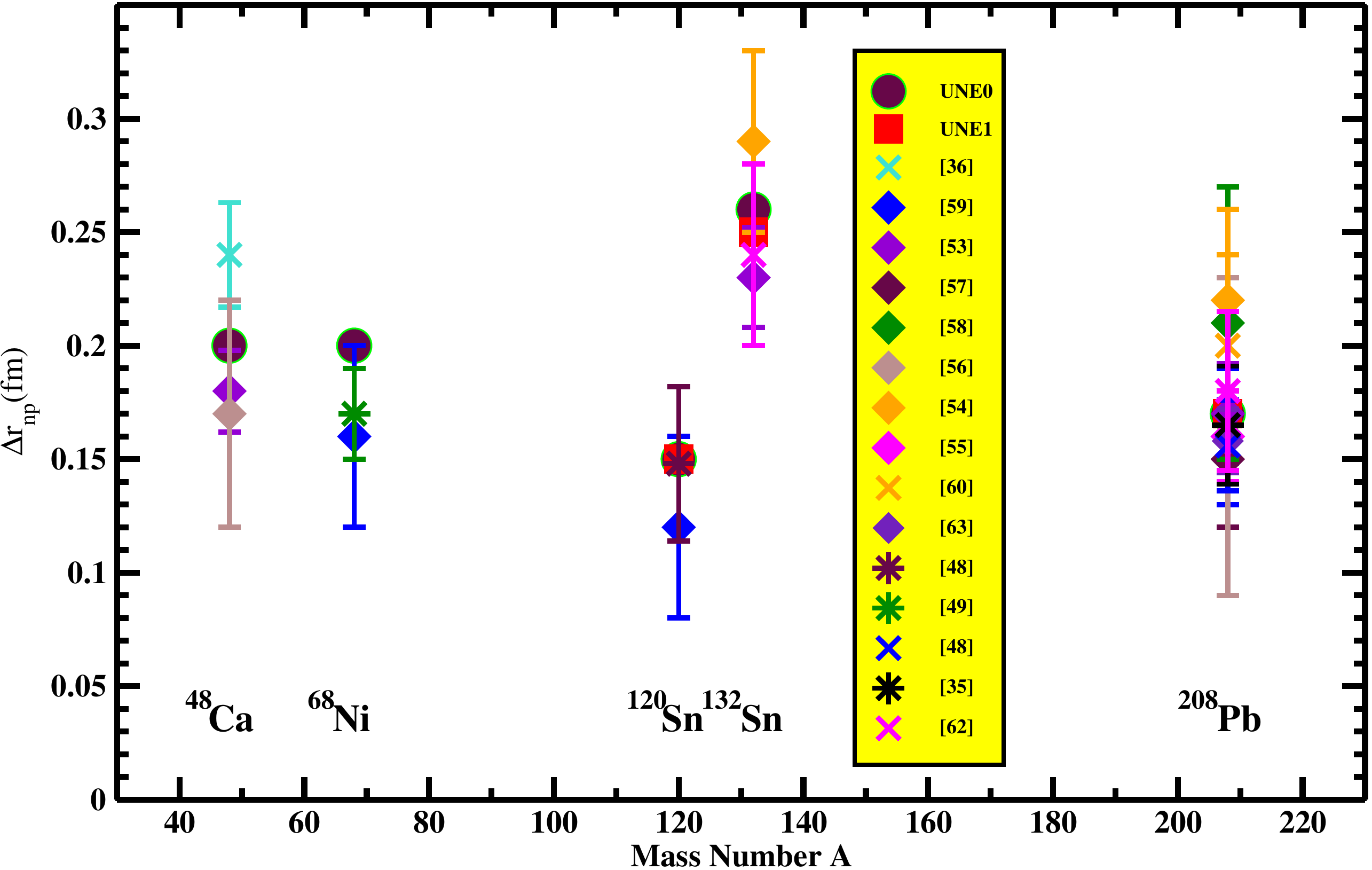}
\caption{\label{fig9}(color online)  Neutron skin thickness $\Delta r _{np} = r_{n}- r_{p}$ in fm calculated theoretically
for doubly magic nuclei. The theoretical estimates are computed from Axially deformed solution of the
Skyrme-Hartree-Fock-Bogoliubov equations using
the  harmonic oscillator basis  (HFB+HO), based on Energy Density Functional (EDF) parameterization
UNEDF0 and UNEDF1. Data is also listed from the different experimental and theoretical results.}
\end{figure}

In lower panel of Fig. (\ref{fig7}), we present the neutron skin thickness ${\Delta r_{np} = r_n - r_p}$ in fm, plotted as a function of atomic number Z,
for even-even isotones of neutron shell closure at N = 28 for $^{40}$Mg, $^{42}$Si, $^{44}$S, $^{46}$Ar, $^{48}$Ca,  $^{50}$Ti and $^{52}$Cr.
The recent observation of $^{40}$Mg \cite{Baumann2007} provides a significant advancement in our understanding of where the neutron drip line 
is likely to go for nuclei with atomic number 12. In this isotopic chain we get maximum value of ${\Delta r_{np}}$ ranging as 0.65-0.76 fm for
$^{40}$Mg and $^{42}$Si. 
The discovery of $^{42}$Si \cite{Fridmann2005} indicates the strong evidence for a well-developed proton sub-shell closure at Z = 14
(14 protons), the near degeneracy of two different (s$_{1/2}$ and d$_{3/2}$)
proton orbits in the vicinity of $^{42}$Si nucleus. Therefore $^{42}$Si nucleus has become the focus of particular interest in
discussions of nuclear shell structure at the neutron drip line.\\
In Fig. (\ref{fig8}), we present the theoretically computed results of neutron skin thickness ${\Delta r_{np}}$ for 
the chain of isotones for neutron numbers at magic number 50 (upper panel) and 82 (lower panel). The magnitude of ${\Delta r_{np}}$ is being obtained as 
0.021 fm in $^{94}$Ru and 0.36 fm in $^{78}$Ni from HFB+HO model. 
The  theoretical value of ${\Delta r_{np}}$ in  doubly magic $^{132}$Sn nucleus  is 0.26 fm, which is comparable with recent experimental 
extractions of ${\Delta r_{np}}$  $0.23\pm0.02$ fm \cite{Piekarewicz2012} and $0.29\pm0.04$ fm \cite{Chen2005}. 
Lower panel of Fig. (\ref{fig8}) presents the variation in ${\Delta r_{np}}$
for isotones from atomic number 46 ($^{128}$Pd) to 68 ($^{150}$Er) and, the magnitude of skin thickness varies as  
0.005 fm in $^{150}$Er - 0.32 fm in $^{128}$Er and  0.01 fm in $^{150}$Er - 0.33 fm in $^{128}$Pd, corresponding to HFB+HO parameterization (UNEDF0). 
The values of ${\Delta r_{np}}$ decreases as the atomic number Z is increasing in the chain of isotones shown in Fig. (\ref{fig8})
At last, in Fig. (\ref{fig9}), we present our theoretical results of neutron skin thickness for neutron-rich doubly-magic nuclei $^{48}$Ca,
 $^{132}$Sn, $^{208}$Pb and $^{68}$Ni, $^{120}$Sn the computed results are compared to the recently available experimental data \cite{Hashimoto2015,Tamii2011,Rossi2013,Klos2007,Friedman2012,Tarbert2014,Zenihiro2010,Klimkiewicz2007} 

 extractions \cite{Piekarewicz2012,Chen2005,Maza2015,Brown2007,Dong2018,Maza2013,Mahzoon2017}.  There are many theoretical and experimental investigations focused on  $^{208}Pb$, $^{132}Sn$, $^{120}Sn$, $^{68}Ni$ and $^{48}Ca$  nuclei,
 which have well understood nuclear structure due to their closed protons and neutrons shells at the magic numbers.
A recent reviews on experimental measurements of  $\Delta r _{np}$ 
in $^{208}Pb$ \cite{Tarbert2014}, suggest that its values ranges from 0.15$\pm$0.03 fm to 0.22$\pm$0.04 fm \cite{Tarbert2014}, with the analysis  of coherent pion photo-production and pion scattering, respectively.  Whereas our theoretical results for 
$\Delta r _{np} = 0.17$ fm in $^{208}Pb$ are reasonable well within the experimental measurements.  

In Fig. (\ref{fig9}) , we have also presented and compared our results of  $\Delta r _{np}$ for $^{132}Sn$, $^{120}Sn$, $^{68}Ni$ and $^{48}Ca$  nuclei. It can be seen clearly that the theoretically estimates
 from HFB+HO model (with both UNEDF0 and UNEDF1 parameterizations) are nicely matching with the recently available experimental data. UNEDF0 parameterization
 are giving very good results of neutron skin thickness while UNEDF0 parameterization are producing even better results of skin thickness for the doubly magic nuclei with higher 
 mass number as we can see in the Fig. For lower mass number range , the result of UNEDF0 parameterization is overestimated by
 approximately 0.03 fm (when compared with experimental data) and almost matching (when compared with experimental data) for $^{48}$Ca. 
 Also for $^{68}$Ni, our result calculated with UNEDF0 parameterization is overestimated by approximately 0.09 fm and 0.1 fm (when both result are compared with the experimental data taken 
 from. The results that we have computed here are taken from  Axially deformed solution of the
Skyrme-Hartree-Fock-Bogoliubov equations using
the  harmonic oscillator model (HFB+HO) based on Energy Density Functional (EDF) parameterizations
UNEDF0 and UNEDF1.

\section{Conclusions}
By employing Nuclear Density Functionals based on HFB+HO model 
 parameterizations (UNEDF0 and UNEDF1) on  various nuclei, we have reached to some striking conclusions, which are discussed in this section.

 The theoretically computed results with NDF parameterizations (UNEDF0 and UNEDF1) for nuclear root-mean-square radii R$_{rms}$, charge radii R$_{ch}$
 and neutron skin thickness ${\Delta r_{np}}$ defined in Eq. (\ref{eq:rnp}) are presented and compared with recent available experimental data. It can be extracted from Fig. (\ref{fig9}) that the ratio Z/N $\approx$ 0.7 
in doubly magic nuclei $^{48}$Ca,  $^{120}$Sn and $^{208}$Pb and the value of  ${\Delta r_{np}}$ lies in close order of 0.16-0.19 fm, 
 whereas for the ratio Z/N $\approx$ 0.5-0.6 in  $^{42}$Si, $^{44}$S and $^{132}$Sn, the value of  ${\Delta r_{np}}$ is more than 0.25 fm. 
 This observation establishes the relationship of ratio Z/N with  ${\Delta r_{np}}$ in the doubly magic and 
 neutron rich nuclei indicating shell closures in the recent investigations \cite{Fridmann2005} in $^{42}$Si nucleus.
 The charge radii for Si, S, Ar and Ca isotopes is observed to be minimum at neutron number N =14. The small magnitudes of R$_c$ in $^{28}$Si, $^{30}$S, $^{32}$Ar and  $^{34}$Ca nuclei suggests N = 14
as a new magic number.
 The theoretically computed results are reasonably reproducing the values for $\Delta r_{np}$ in $^{208}Pb$, $^{120,132}Sn$, 
 and $^{68}Ni$ nuclei are in the ranges 0.13 - 0.19 fm, 0.12 - 0.16 fm, and 0.15 - 0.19 fm, respectively from ref. \cite{Maza2015}, 
 whereas the in case of $^{48}$Ca, our results overestimated by very small value of  approximately 0.01 fm only. These results confirms the validity of skyrme-interactions with zero range pairing spin interactions for the exotic nuclei. 
\section{Acknowledgements} Author(s) would like to thank Himachal Pradesh University for providing computational facilities, DST-
INSPIRE for providing financial assistance (Junior Research Fellowship) and anonymous 
refree(s) for  extremely thorough inspection of the manuscript and helpful comments.

\begin{thebibliography}{10}
\expandafter\ifx\csname url\endcsname\relax
  \def\url#1{\texttt{#1}}\fi
\expandafter\ifx\csname urlprefix\endcsname\relax\def\urlprefix{URL }\fi
\expandafter\ifx\csname href\endcsname\relax
  \def\href#1#2{#2} \def\path#1{#1}\fi

\bibitem{Lunderberg2016}
E.~Lunderberg, et ~al.
\newblock In-beam $\ensuremath{\gamma}$-ray spectroscopy of
  $^{38--42}\mathrm{S}$.
\newblock {\em Phys. Rev. C}, 94:064327, Dec 2016.

\bibitem{Gade2009}
A.~Gade, et ~al.
\newblock In-beam $\ensuremath{\gamma}$-ray spectroscopy of very neutron-rich
  nuclei: Excited states in $^{46}\mathbf{S}$ and $^{48}\mathrm{Ar}$.
\newblock {\em Phys. Rev. Lett.}, 102:182502, May 2009.

\bibitem{Coraggio2009}
L.~Coraggio, et ~al.
\newblock Spectroscopic study of neutron-rich calcium isotopes with a realistic
  shell-model interaction.
\newblock {\em Phys. Rev. C}, 80:044311, Oct 2009.

\bibitem{Papakonstantinou2018}
Panagiota Papakonstantinou, et ~al.
\newblock Density dependence of the nuclear energy-density functional.
\newblock {\em Phys. Rev. C}, 97:014312, Jan 2018.

\bibitem{Niksic2008}
T.~Nik\ifmmode \check{s}\else \v{s}\fi{}i\ifmmode~\acute{c}\else \'{c}\fi{},
  et ~al.
\newblock Relativistic nuclear energy density functionals: Adjusting parameters
  to binding energies.
\newblock {\em Phys. Rev. C}, 78:034318, Sep 2008.

\bibitem{Bhuyan2018}
M.~Bhuyan, et ~al.
\newblock Surface properties of neutron-rich exotic nuclei within relativistic
  mean field formalisms.
\newblock {\em Phys. Rev. C}, 97:024322, Feb 2018.
\bibitem{Gade2006}
A.~Gade, et ~al.
\newblock Cross-shell excitation in two-proton knockout: Structure of
  $^{52}\mathrm{Ca}$.
\newblock {\em Phys. Rev. C}, 74:021302, Aug 2006.

\bibitem{Utsuno2012}
Yutaka Utsuno, et ~al.
\newblock Shape transitions in exotic si and s isotopes and tensor-force-driven
  jahn-teller effect.
\newblock {\em Phys. Rev. C}, 86:051301, Nov 2012.

\bibitem{Tostevin2013}
J.~A. Tostevin, et ~al.
\newblock Two-proton removal from ${}^{44}$s and the structure of ${}^{42}$si.
\newblock {\em Phys. Rev. C}, 87:027601, Feb 2013.

\bibitem{Bhattacharyya2008}
S.~Bhattacharyya, et ~al.
\newblock Structure of neutron-rich ar isotopes beyond $n=28$.
\newblock {\em Phys. Rev. Lett.}, 101:032501, Jul 2008.

\bibitem{Utsuno2015}
Yutaka Utsuno, et ~al.
\newblock Nature of isomerism in exotic sulfur isotopes.
\newblock {\em Phys. Rev. Lett.}, 114:032501, Jan 2015.

\bibitem{Stroberg2014}
S.~R. Stroberg, et ~al.
\newblock Single-particle structure of silicon isotopes approaching
  $^{42}\mathrm{Si}$.
\newblock {\em Phys. Rev. C}, 90:034301, Sep 2014.
\bibitem{Sorlin1993}
O.~Sorlin, et ~al.
\newblock Decay properties of exotic n\ensuremath{\simeq}28 s and cl nuclei and
  the $^{48}\mathrm{Ca}$${/}^{46}$ca abundance ratio.
\newblock {\em Phys. Rev. C}, 47:2941--2953, Jun 1993.

\bibitem{Schneider1994}
R~Schneider, et ~al.
\newblock Production and identification of 100 sn.
\newblock {\em Zeitschrift f{\"u}r Physik A Hadrons and Nuclei},
  348(4):241--242, 1994.

\bibitem{Leander1975}
G~Leander and SE~Larsson.
\newblock Potential-energy surfaces for the doubly even n= z nuclei.
\newblock {\em Nuclear Physics A}, 239(1):93--113, 1975.

\bibitem{Jaqaman1984}
HR~Jaqaman and L~Zamick.
\newblock High multipole moments in nuclei.
\newblock {\em Physical Review C}, 30(5):1719, 1984.

\bibitem{Flocard1984}
H~Flocard, et ~al.
\newblock Configuration space, cranked hartree-fock calculations for the nuclei
  16o, 24mg and 32s.
\newblock {\em Progress of Theoretical Physics}, 72(5):1000--1016, 1984.

\bibitem{Girod1983}
M.~Girod and B.~Grammaticos.
\newblock {\em Phys. Rev. C}, $\mathbf{27}$:2317, 1983.

\bibitem{Furnstahl1987}
RJ~Furnstahl.
\newblock Rj furnstahl, ce price, and ge walker, phys. rev. c 36, 2590 (1987).
\newblock {\em Phys. Rev. C}, 36:2590, 1987.

\bibitem{Fink1989}
J~Fink, et ~al.
\newblock Systematic study of potential energy surfaces of light nuclei in
  relativistic hartree calculations.
\newblock {\em Physics Letters B}, 218(3):277--282, 1989.

\bibitem{Patra1993}
SK~Patra and CR~Praharaj.
\newblock Shapes of n= z nuclei in the mass a= 20--48 region.
\newblock {\em Nuclear Physics A}, 565(2):442--454, 1993.

\bibitem{Zhang1994}
Jian-Kang Zhang and DS~Onley.
\newblock Systematic relativistic hartree-fock calculation of deformed nuclei
  in s-d shell.
\newblock {\em Physical Review C}, 49(2):762, 1994.
\bibitem{Carchidi1986}
M~Carchidi, BH~Wildenthal, and B~Alex Brown.
\newblock Quadrupole moments of sd-shell nuclei.
\newblock {\em Physical Review C}, 34(6):2280, 1986.
\bibitem{Zhang1994(1)}
J~Zhang, WDM Rae, and AC~Merchant.
\newblock Systematics of some 3-dimensional $\alpha$-cluster configurations in
  4n nuclei from 16o to 44ti.
\newblock {\em Nuclear Physics A}, 575(1):61--71, 1994.
\bibitem{Yang2018}
Junjie Yang and J.~Piekarewicz.
\newblock Difference in proton radii of mirror nuclei as a possible surrogate
  for the neutron skin.
\newblock {\em Phys. Rev. C}, 97:014314, Jan 2018.

\bibitem{Brown2017}
B.~Alex Brown.
\newblock Mirror charge radii and the neutron equation of state.
\newblock {\em Phys. Rev. Lett.}, 119:122502, Sep 2017.

\bibitem{Brown2009}
B.~A. Brown, et ~al.
\newblock Calculations of the neutron skin and its effect in atomic parity
  violation.
\newblock {\em Phys. Rev. C}, 79:035501, Mar 2009.
\bibitem{Centelles2009}
M~Centelles, et ~al.
\newblock Nuclear symmetry energy probed by neutron skin thickness of nuclei.
\newblock {\em Physical review letters}, 102(12):122502, 2009.
\bibitem{Korte2012}
M.~Kortelainen, et ~al.
\newblock {\em Phys. Rev, C}, $\mathbf{85}$:024304, 2012.
\bibitem{Stoitsov2005}
M.V. Stoitsov, et ~al.
\newblock {\em Computer Physics Communications}, $\mathbf{167}$:43--63, 2005.
\bibitem{Kortelainen2010}
M.~Kortelainen, T.~Lesinski, J.~Mor\'e, W.~Nazarewicz, J.~Sarich, N.~Schunck,
  M.~V. Stoitsov, and S.~Wild.
\newblock Nuclear energy density optimization.
\newblock {\em Phys. Rev. C}, 82:024313, Aug 2010
\bibitem{Ring1980}
Peter Ring and Peter Schuck.
\newblock {\em The nuclear many-body problem}.
\newblock Springer Science \& Business Media, 2004.

\bibitem{Bender2003}
Michael Bender, Paul-Henri Heenen, and Paul-Gerhard Reinhard.
\newblock Self-consistent mean-field models for nuclear structure.
\newblock {\em Reviews of Modern Physics}, 75(1):121, 2003.

\bibitem{Skyrme1959}
THR Skyrme.
\newblock Thr skyrme, nucl. phys. 9, 615 (1959).
\newblock {\em Nucl. Phys.}, 9:615, 1959.
\bibitem{Lipkin1960}
H.J. Lipkin.
\newblock {\em Ann. Phys.}, $\mathbf{9}$:272, 1960.
\bibitem{Dobaczewski1993}
J.~Dobaczewski and W.~Nazarewicz.
\newblock {\em Phys. Rev, C}, $\mathbf{47}$:2418, 1993.
\bibitem{Stoitsov2013}
M.V. Stoitsov, N.~Schunck, M.~Kortelainen, N.~Michel, H.~Nam, E~Olsen,
  J.~Sarich, and S.~Wild.
\newblock {\em Computer Physics Communications}, $\mathbf{184}$:1592--1604,
  2013.
  
\bibitem{Garcia2016}
R.~F. Garcia~Ruiz,et ~al,
  \href{http://dx.doi.org/10.1038/nphys3645}{Unexpectedly large charge radii of
  neutron-rich calcium isotopes}, Nature Physics 12 (2016) 594--598.
\newblock \href {http://dx.doi.org/10.1038/nphys3645}
  {\path{doi:10.1038/nphys3645}}.
\newline\urlprefix\url{http://dx.doi.org/10.1038/nphys3645}
\bibitem{Bastin2007}
B.~Bastin and S.~Gr\'evy.
\newblock Collapse of the $n=28$ shell closure in $^{42}\mathrm{S}\mathrm{i}$.
\newblock {\em Phys. Rev. Lett.}, $\mathbf{99}$:022503, Jul 2007.

\bibitem{Reinhard0000}
P.~G. Reinhard, W.~Nazarewicz, {Information content of a new observable: The
  case of the nuclear neutron skin}, Physical Review C - Nuclear Physics
  81~(5).
\newblock \href {http://arxiv.org/abs/1002.4140} {\path{arXiv:1002.4140}},
  \href {http://dx.doi.org/10.1103/PhysRevC.81.051303}
  {\path{doi:10.1103/PhysRevC.81.051303}}.


\bibitem{Reinhard2016}
P.-G. Reinhard and W.~Nazarewicz.
\newblock Nuclear charge and neutron radii and nuclear matter: trend analysis.
\newblock {\em arXiv:1601.06324v1 [nucl-th]}, Jan 2016.

\bibitem{Hagen2015}
G.~Hagen, et ~al.
\newblock Charge, neutron, and weak size of the atomic nucleus.
\newblock {\em Nat Phys, Advance Online Publication/arXiv:1509.07169
  [nucl-th]}, 2015.

\bibitem{Angeli2013}
I.~Angeli and K.P. Marinova.
\newblock Table of experimental nuclear ground state charge radii: An update.
\newblock {\em Atomic Data and Nuclear Data Tables}, $\mathbf{99}$(1):69--95,
  2013.
\bibitem{Agrawal2006}
B.~K. Agrawal, Shashi~K. Dhiman, and Raj Kumar.
\newblock Exploring the extended density-dependent skyrme effective forces for
  normal and isospin-rich nuclei to neutron stars.
\newblock {\em Phys. Rev. C}, $\mathbf{73}$:034319, Mar 2006.

\bibitem{Lattimer2013}
James~M. Lattimer and Yeunhwan Lim.
\newblock Constraining the symmetry parameters of the nuclear interaction.
\newblock {\em The Astrophysical Journal}, $\mathbf{771}$(1):51, 2013.


\bibitem{Paar}
T.~Niksic, N.~Paar, D.~Vretenar, and P.~Ring.
\newblock Dirhb—a relativistic self-consistent mean-field framework for
  atomic nuclei.
\newblock {\em Computer Physics Communications}, $\mathbf{185}$(6):1808--1821,
  2014.

\bibitem{Tamii2011}
A.~Tamii, et ~al.
\newblock Complete electric dipole response and the neutron skin in
  $^{208}\mathrm{Pb}$.
\newblock {\em Phys. Rev. Lett.}, $\mathbf{107}$:062502, Aug 2011.


\bibitem{Hashimoto2015}
T.~Hashimoto, et ~al.
\newblock Dipole polarizability of $^{120}\mathbf{Sn}$ and nuclear energy
  density functionals.
\newblock {\em arXiv:1503.08321 [nucl-ex]}, 2015.

\bibitem{Rossi2013}
D.~M. Rossi, et ~al.
\newblock Measurement of the dipole polarizability of the unstable neutron-rich
  nucleus $^{68}\mathrm{Ni}$.
\newblock {\em Phys. Rev. Lett.}, $\mathbf{111}$:242503, Dec 2013.


\bibitem{Nazarewicz2014}
W.~Nazarewicz, et ~al.
\newblock Symmetry energy in nuclear density functional theory.
\newblock {\em The European Physical Journal A $\mathbf{50}$:20}, Feb 2014.


\bibitem{Baumann2007}
T.~Baumann, et ~al.
\newblock Discovery of $^{40}\mathrm{Mg}$ and $^{42}\mathrm{Al}$ suggests
  neutron drip-line slant towards heavier isotopes.
\newblock {\em Nature}, $\mathbf{449}$:1022--1024, oct 2007.

\bibitem{Fridmann2005}
J.~Fridmann, et ~al.
\newblock Magic nucleus $^{42}\mathrm{Si}$.
\newblock {\em Nature}, $\mathbf{435}$:922--924, June 2005.

\bibitem{Piekarewicz2012}
J.~Piekarewicz, B.~K. Agrawal, G.~Col\`o, W.~Nazarewicz, N.~Paar, P.-G.
  Reinhard, X.~Roca-Maza, and D.~Vretenar.
\newblock Electric dipole polarizability and the neutron skin.
\newblock {\em Phys. Rev. C}, $\mathbf{85}$:041302, Apr 2012.

\bibitem{Chen2005}
Lie-Wen Chen, Che~Ming Ko, and Bao-An Li.
\newblock Nuclear matter symmetry energy and the neutron skin thickness of
  heavy nuclei.
\newblock {\em Phys. Rev. C}, $\mathbf{72}$:064309, Dec 2005.


\bibitem{Klos2007}
B.~K\l{}os, et ~al.
\newblock Neutron density distributions from antiprotonic $^{208}\mathrm{Pb}$
  and $^{209}\mathrm{Bi}$ atoms.
\newblock {\em Phys. Rev. C}, $\mathbf{76}$:014311, Jul 2007.

\bibitem{Friedman2012}
E.~Friedman.
\newblock Neutron skins of $^{208}\mathrm{Pb}$ and $^{48}\mathrm{Ca}$ from
  pionic probes.
\newblock {\em Nuclear Physics A}, $\mathbf{896}$:46 -- 52, 2012.

\bibitem{Tarbert2014}
C.~M. Tarbert, et ~al.
\newblock Neutron skin of $^{208}\mathrm{Pb}$ from coherent pion
  photoproduction.
\newblock {\em Phys. Rev. Lett.}, $\mathbf{112}$:242502, Jun 2014.

\bibitem{Zenihiro2010}
J.~Zenihiro, et ~al.
\newblock Neutron density distributions of $^{204,206,208}\mathrm{Pb}$ deduced
  via proton elastic scattering at ${E}_{p}=295$ mev.
\newblock {\em Phys. Rev. C}, $\mathbf{82}$:044611, Oct 2010.

\bibitem{Maza2015}
X.~Roca-Maza, et ~al.
\newblock Neutron skin thickness from the measured electric dipole polarizability in $^{68}\text{Ni}$, $^{120}\text{Sn}$, and $^{208}\text{Pb}$.
\newblock {\em Phys. Rev. C}, $\mathbf{92}$:064304, Dec 2015.


\bibitem{Brown2007}
B.~Alex Brown, et ~al.
\newblock Neutron skin deduced from antiprotonic atom data.
\newblock {\em Phys. Rev. C}, $\mathbf{76}$:034305, Sep 2007.


\bibitem{Wang2012}
M.Wang, et ~al.
\newblock The ame2012 atomic mass evaluation.
\newblock {\em Chinese Physics C}, $\mathbf{36}$:1603--2014, 2012.

\bibitem{Klimkiewicz2007}
A.~Klimkiewicz, et ~al,
  \href{https://link.aps.org/doi/10.1103/PhysRevC.76.051603}{Nuclear symmetry
  energy and neutron skins derived from pygmy dipole resonances}, Phys. Rev. C
  76 (2007) 051603.
\newblock \href {http://dx.doi.org/10.1103/PhysRevC.76.051603}
  {\path{doi:10.1103/PhysRevC.76.051603}}.
\newline\urlprefix\url{https://link.aps.org/doi/10.1103/PhysRevC.76.051603}

\bibitem{Dong2018}
J.~M. Dong, L.~J. Wang, W.~Zuo, J.~Z. Gu,
  \href{https://link.aps.org/doi/10.1103/PhysRevC.97.034318}{{Constraints on
  Coulomb energy, neutron skin thickness in $^{208}$Pb, and
  symmetry energy}}, Physical Review C 97~(3) (2018) 034318.
\newblock \href {http://dx.doi.org/10.1103/PhysRevC.97.034318}
  {\path{doi:10.1103/PhysRevC.97.034318}}.
\newline\urlprefix\url{https://link.aps.org/doi/10.1103/PhysRevC.97.034318}

\bibitem{Maza2013}
X.~Roca-Maza, et ~al, {Electric dipole
  polarizability in $^{208}$Pb: Insights from the droplet model}, Physical Review C
  - Nuclear Physics 88~(2) (2013) 1--7.
\newblock \href {http://arxiv.org/abs/1307.4806} {\path{arXiv:1307.4806}},
  \href {http://dx.doi.org/10.1103/PhysRevC.88.024316}
  {\path{doi:10.1103/PhysRevC.88.024316}}.

 \bibitem{Mahzoon2017}
M.~H. Mahzoon, M.~C. Atkinson, R.~J. Charity, W.~H. Dickhoff,
  \href{https://link.aps.org/doi/10.1103/PhysRevLett.119.222503}{Neutron skin
  thickness of $^{48}$Ca from a nonlocal dispersive optical-model
  analysis}, Phys. Rev. Lett. 119 (2017) 222503.
\newblock \href {http://dx.doi.org/10.1103/PhysRevLett.119.222503}
  {\path{doi:10.1103/PhysRevLett.119.222503}}.
\newline\urlprefix\url{https://link.aps.org/doi/10.1103/PhysRevLett.119.222503}
\end{thebibliography}

\clearpage

\end{document}